# CESR's POSITRON SOURCE

## A. Mikhailichenko

*Wilson Laboratory, Cornell University, Ithaca, NY 14853*

**Abstract**

Some details of the new positron converter and power supply are described. This converter unit so far has doubled the amount of positrons accelerated in CESR.

## 1. INTRODUCTION

New positron converter was installed and tested at CESR complex [1].
This accomplished the stage, which deals with the collection optics; i.e. elements located right after the target.
It also includes the new power supply construction and testing.

At the CESR complex a bi-layer solenoid was used as a short focusing lens, located right after the target. This lens provided a Quarter Wave Transformation (QWT) i.e. the target was located on a focal plane of this lens.
This type of focusing system was originated in [2].

The target was located at the end of a paddle-type holder in front of this bi-layer coil.

The requirement for positron accumulation in CESR at the minimal level 100 *mA/min* yielded a necessity for a new positron conversion unit.

In this unit all accumulated knowledge about the system was to be used for reaching this goal and pave the road for further improvements.

We represent here a general description of what was done on the way. So far it doubled the rate of positron accumulation at CESR.

The device for positron collection can be considered as a model for future positron collection optics for LC.

---

# 2. BASIC CONCEPTS OF NEW COLLECTION SYSTEM

Four basic principles of collection system are:

**shorter focal distance** of the lens with its closer positioning to the target;
**possibility of alignment** of converter unit with respect to accelerating structure;
**symmetrical allocation for input wires**;
**symmetry of the surroundings**.

*First principle*

The shorter focal length allows smaller beam sizes.
For some extent if there is no limitation in transverse size of the lens, this can allow an increase of capturing angle.

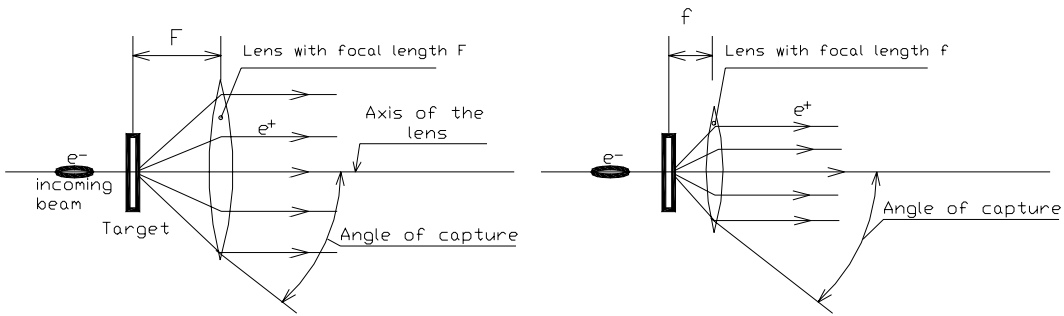

The geometry of capturing. Target located at the distance F (or f) -the focal distance of the lens.
Shorter focal distance–smaller the beam size, as the divergence of the beam remains the same.

*Second principle*
If the hot spot is off-axis with respect to the focal point of the lens, the secondary beam receives an angle $\alpha \cong \dfrac{x}{F}$, where *x* is a misalignment.

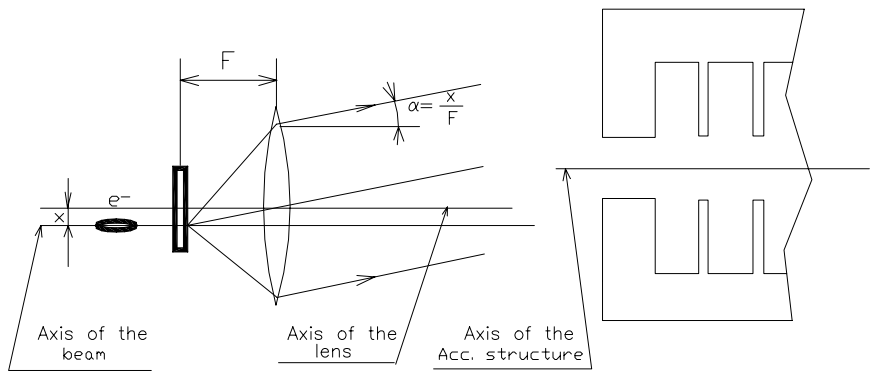

The beam line and the lens axis have a shift *x*. All three axes must be congruent.

The focal distance *F* for the solenoidal type of lens used is given by

$$F = \frac{4(HR)^2}{\int H^2(s)ds},$$

where (*HR*) is a magnet rigidity of the particle, *pc=300(HR)*, *H(s)* is a longitudinal field distribution on the axis, *s* is a longitudinal coordinate.

One can see, that the focal distance has a *quadratic* dependence on energy (and on feeding current *I*, $H \propto I$ ).

The angle of the plane rotation is $\theta = \int H(s)ds/2(HR)$. These formulas can be easily obtained after the field consideration in 3D, as

$$H_r(r,s) \cong -\frac{r}{2}\frac{\partial H(s)}{\partial s} + \frac{r^3}{16}\frac{\partial^3 H(s)}{\partial s^3} - ... , \quad H_s(r,s) \cong H(s) - \frac{r^2}{4}\frac{\partial^2 H(s)}{\partial s^2} - ... .$$

The angular kick $\alpha$ for fixed mismatch has a quadratic dependence on energy and feeding current in a solenoidal lens as

$$\alpha \cong \frac{x}{F} \cong \frac{x \cdot \int H^2(s)ds}{4(B\rho)^2},$$

where *x* –is the transverse offset of the primary beam with respect to the axis of solenoid.

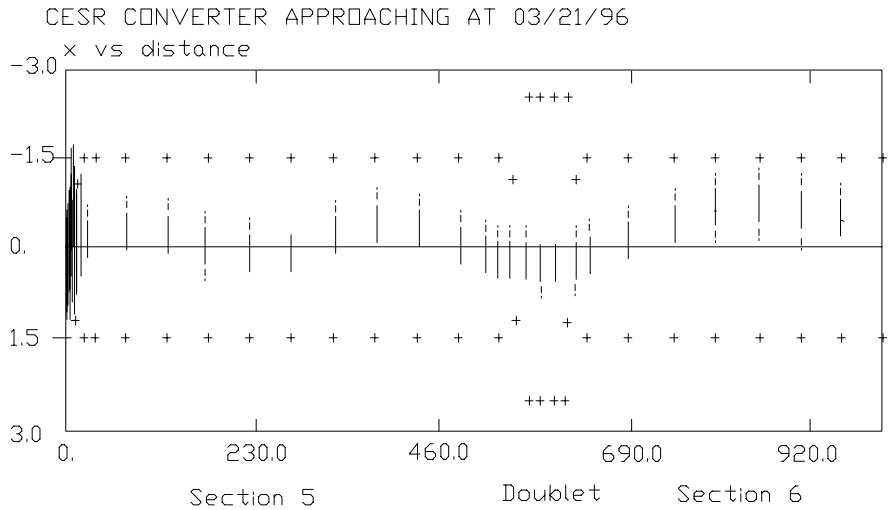

Transverse misalignment of the beam trajectory in sections 5 and 6. Energy of the positrons is 5*MeV*. Current in solenoidal lens ~4*kA*, misalignment $\Delta y = 3mm$. Crosses mark aperture limits.

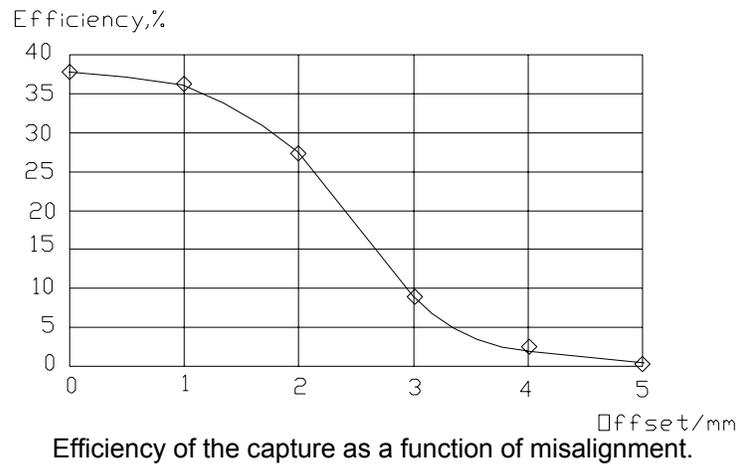
Efficiency of the capture as a function of misalignment.

***Third and fourth principles*** –symmetry in input wires and surroundings are in line of what was just described above.

Any of these asymmetries makes it impossible to increase the focusing without introduction of transverse kick.

So the effect of increasing the capturing efficiency becomes diminished by losses generated by kick.

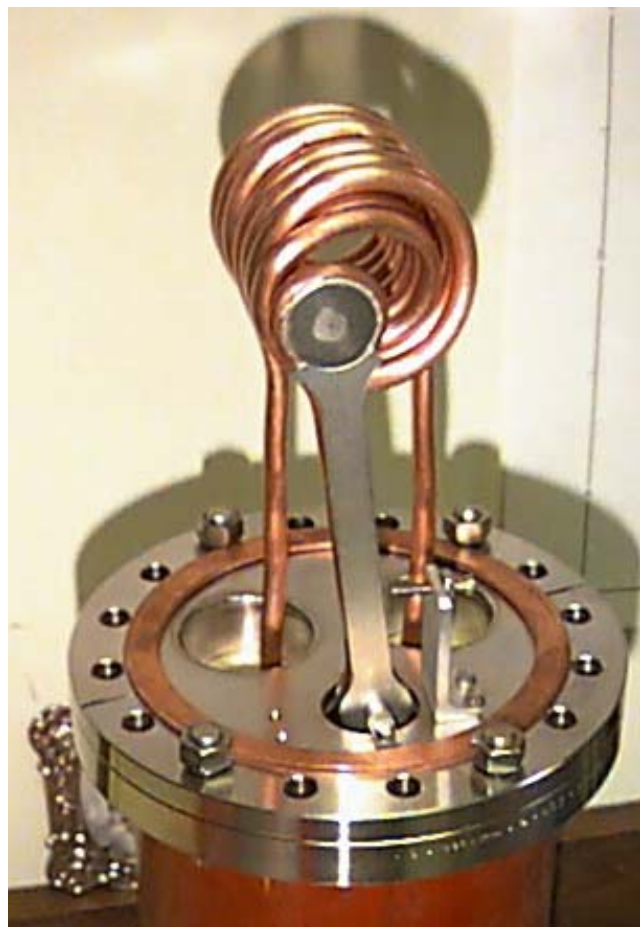
Old pulsed coil

# 3. COLLECTION OPTICS

***Geometry*** of the modified capturing optics is represented in Figure. The focusing coil is working in pulsed mode.

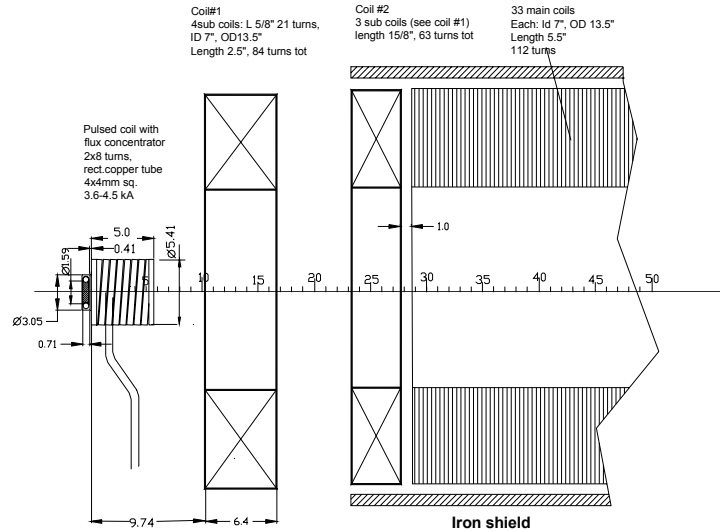

The converter assembly with focusing solenoid. Primary electron beam is coming from the left.

The face plane of the first coil with large diameter is located levelly with the input plane of accelerator structure, section #5.

In the gap between this first coil and main section of solenoid the input waveguide is squeezed. Saying ahead the current in this first coil for better performance must be zero. Right now this coil is fed in series with the main solenoid.

***New target*** made of an alloy of 97% Tungsten, 2.1% Nickel and 0.9% Iron. Radiation lengths for tungsten W is $X_W \cong 8 g/cm^2$, geometric length, corresponding to this radiation length, is $LX_W \cong 0.35 cm$, for Iron $X_{Fe} \sim 13.8 g/cm^2$, $LX_{Fe} \cong 1.75 cm$. So the effective radiation length goes as

$$\frac{1}{LX_{eff}} \cong \frac{W_W}{LX_W} + \frac{W_{Fe}}{LX_{Fe}} \cong \frac{0.97}{3.5} + \frac{0.03}{17.5} \cong \frac{1}{3.58},$$

what gives $LX_{eff} \cong 3.58\ mm$. Optimal thickness of the target has a logarithmic dependence on energy

$$l_{opt} / X_{eff} \cong 1.1 \ln E[GeV] + 3.9.$$

For a 200*MeV* primary electron beam this goes to $l_{opt} \cong 2.12 X_{eff}$, which gives $Ll_{opt} \cong 2.12 X_{eff} \cong 7.59 mm$.

We have chosen, however, the thickness of the target $L \cong 7.063\ mm$, i.e. a little bit less that calculated above.

The reason for this is a weak dependence of positron production versus thickness around optimum. If the target is thinner, there will be fewer problems with heating also.

Another important number to remember is the skin depth of the pulsed field in Tungsten, which is around 1 *mm* for the pulse parameters.

***Numerical calculations*** were carried out with the PARMELA code. The files with initial positron distributions were generated using results [3][3]. Here the transverse distribution of the positrons created by an electron at the hot spot is defined as

$$\frac{d^2N^+}{dxd\theta_x} = \frac{1}{\pi\varepsilon_x^+}\exp\{-\frac{\gamma x^2 + 2\alpha x\theta_x + \beta\theta_x^2}{\varepsilon_x^+}\},$$

where emittance is defined as

$$\varepsilon_x^+ = 2\sqrt{<x^2><\theta_x^2> - <x\theta_x>^2}, \quad \gamma = 2<\theta_x^2>/\varepsilon_x^+, \quad \alpha = -2<x\theta_x>/\varepsilon_x^+, \quad \beta = 2<x^2>/\varepsilon_x^+.$$

Here $x$ measured as a fraction of the radiation length, i.e. $x = 1$ means, that $x$ equals the length, corresponding the radiation length, which is 3.5 *mm* for Tungsten. The brackets mean to average over the ensemble.

Equation $\gamma x^2 + 2\alpha x\theta_x + \beta\theta_x^2 \leq \varepsilon_x^+$ describes the phase ellipse with about 63% of the particles in it.

The values for different energies and for the target having a thickness of $2X_0$ and an incoming beam with energy 200 *MeV* are the following

|  | $E^+$ = 5 MeV | $E^+$ = 10 MeV | $E^+$ = 20 MeV |
|---|---|---|---|
| $<x^2>$ | 0.043 | 0.05 | 0.032 |
| $<\theta_x^2>$ | 0.35 | 0.22 | 0.12 |
| $<x\theta_x>$ | 0.025 | 0.052 | 0.038 |
| $\varepsilon_x^+ cm\cdot rad$ ,(3.4) | 0.084 | 0.085 | 0.034 |

As the current pulse in the lens lasts for about 25 microseconds (see below), the current induced inside the Tungsten target by the pulsed magnetic field penetrates to a depth about 0.8 *mm*. On the other hand, the RMS depth *l* of positron creation $l \cong \frac{<x\theta_x>}{<\theta_x^2>}$ can be estimated as 0.83 *mm* for 10 *MeV* positrons.

The drop of the magnetic field begins at the distance of the order of the coil radius as a result of the counter-current induced in the target. So one can conclude that the most of the particles are created in very low longitudinal magnetic field. This was modeled by the introduction of image currents.

---

[3] [3] V.A. Tajursky, *Calculation of Electron Conversion into Positrons at 0.2—2GeV*, BudkerINP 76-36, Novosibirsk, 1976.

For 10 *MeV* positrons geometrical capture efficiency was found to be 40%.
So $\Delta N_{10}^{+} \cong 0.036$ or 3.6% for 10 *MeV*.

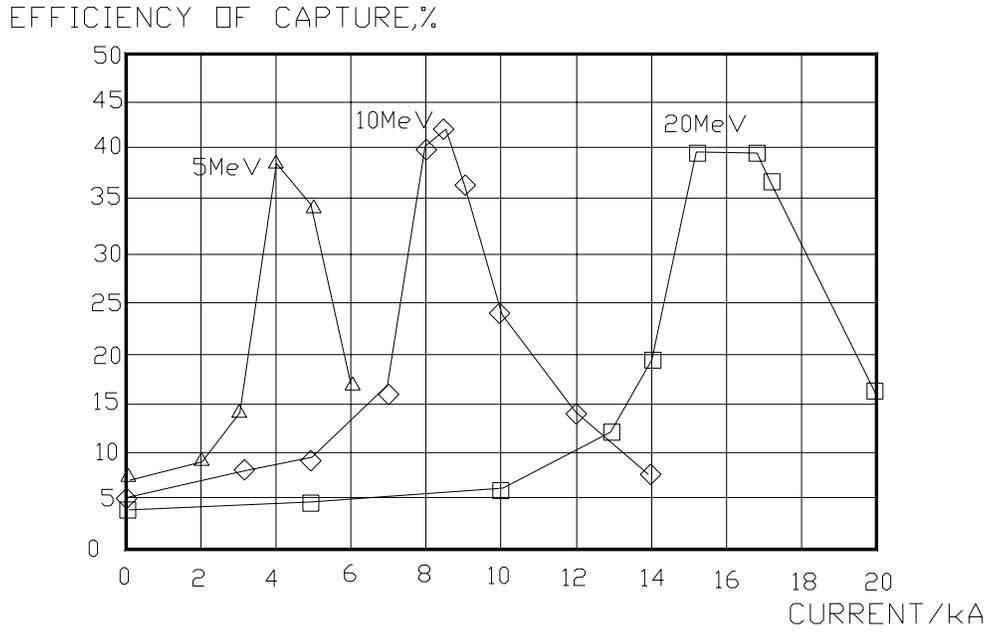

Efficiency of capture for three different values of energy as a function of the feeding current in the pulsed lens.

***Few possibilities*** for short focusing optics were considered.

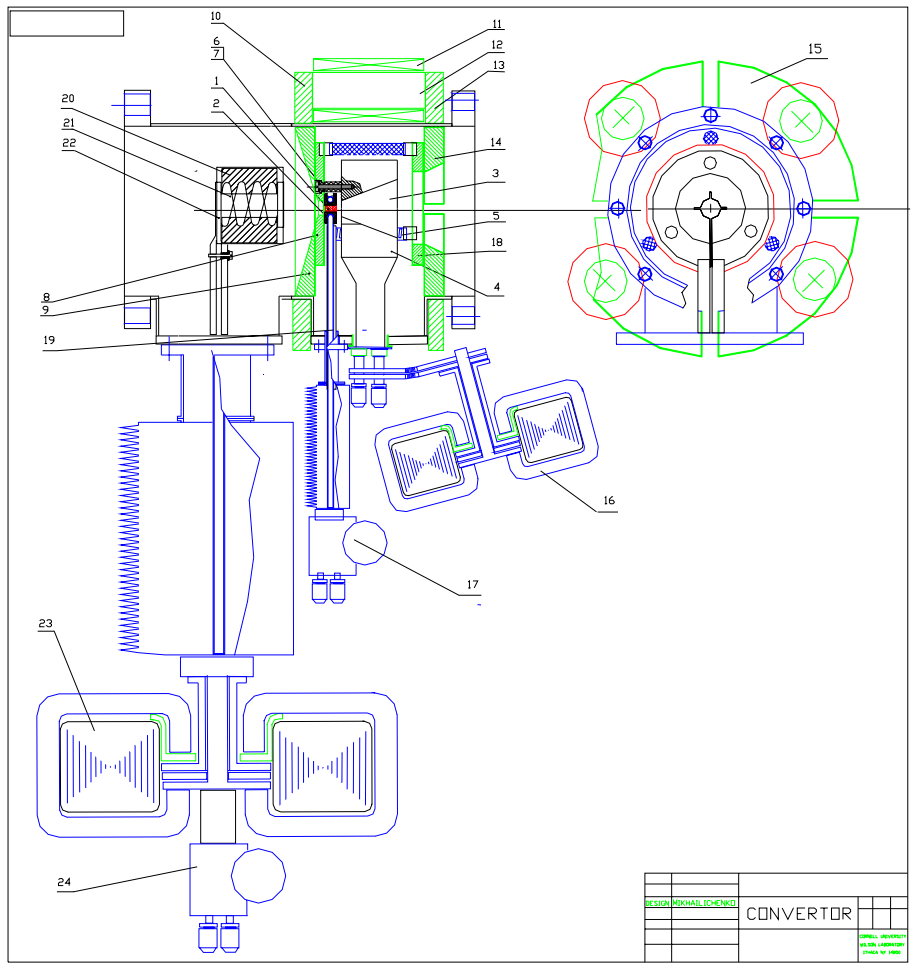

Device represented above uses a single turn loop as a short focusing lens. Feeding transformer located outside close to the housing. Helical quadrupole is shown here as a short focusing element instead of Li lens.

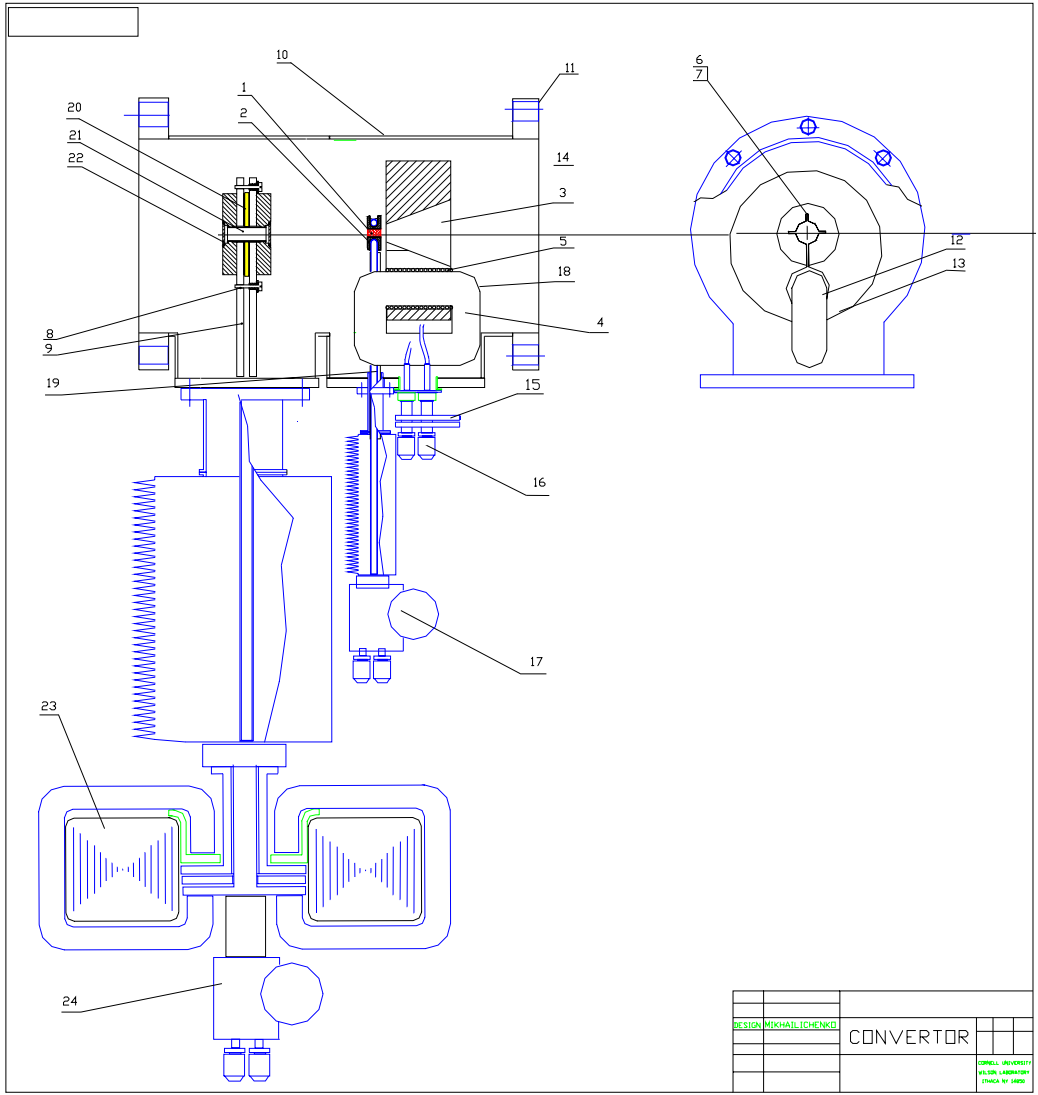

Device shown here uses combined transformer as short focusing lens.

***Reduced version*** with partial flux concentrator was chosen as a basic one. At first stage the focusing element (Li lens or helical quadrupole) is omitted. The flange for further installation is reserved however.

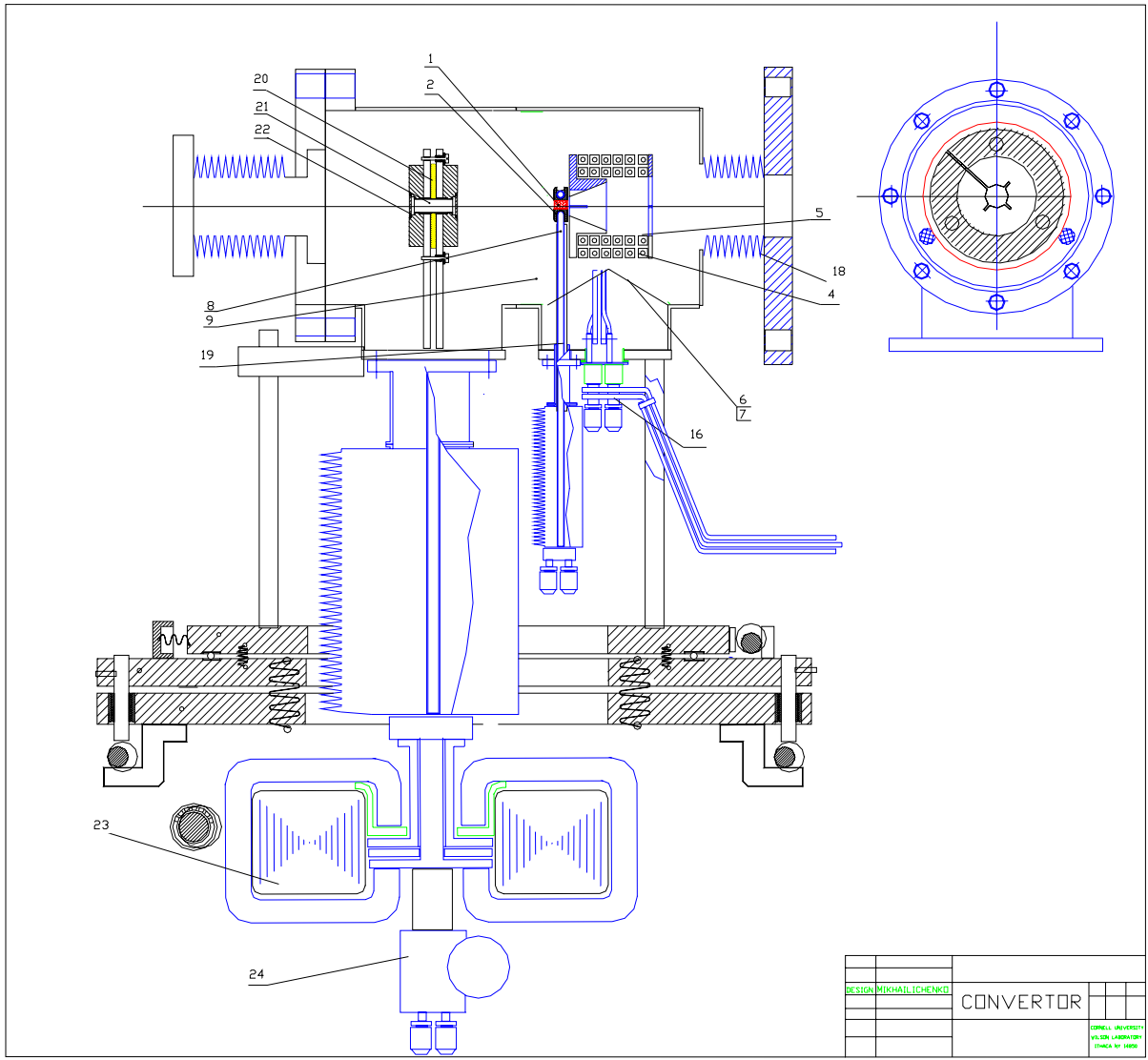

Scheme with partial flux concentrator and Lithium lens. This one became a basic one.

***Coil-Concentrator* is** represented in figure below. It is wound with Oxygen free copper conductor having $4\times4$ *mm$^2$* square cross section with a water hole of 2.5 *mm* in diameter. It has 16 turns total in two layers.

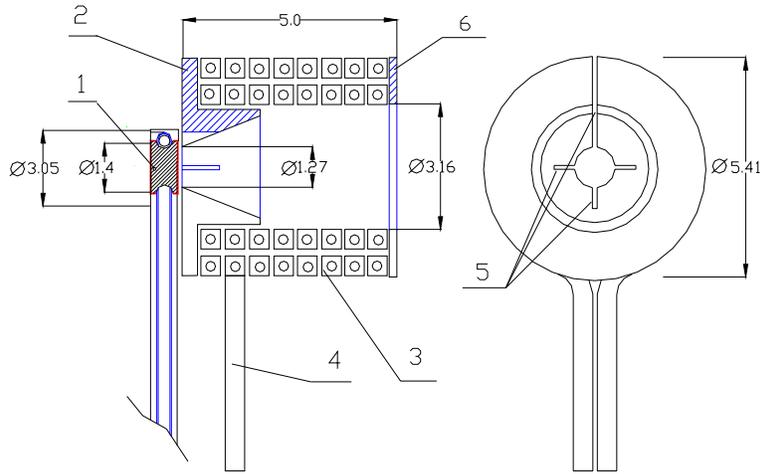

Scaled view on the focusing coil. 1-target, 2-flux concentrator, 3-bilayer solenoid, 4-feeding leads, 5-slots, 6-end plate. Copper conductor has cross section of 4x4 mm$^2$. Dimensions are given in cm.

The input leads are running in the shadow region of the flux concentrator.
For symmetryzation three more semi-slots across the neck were added.

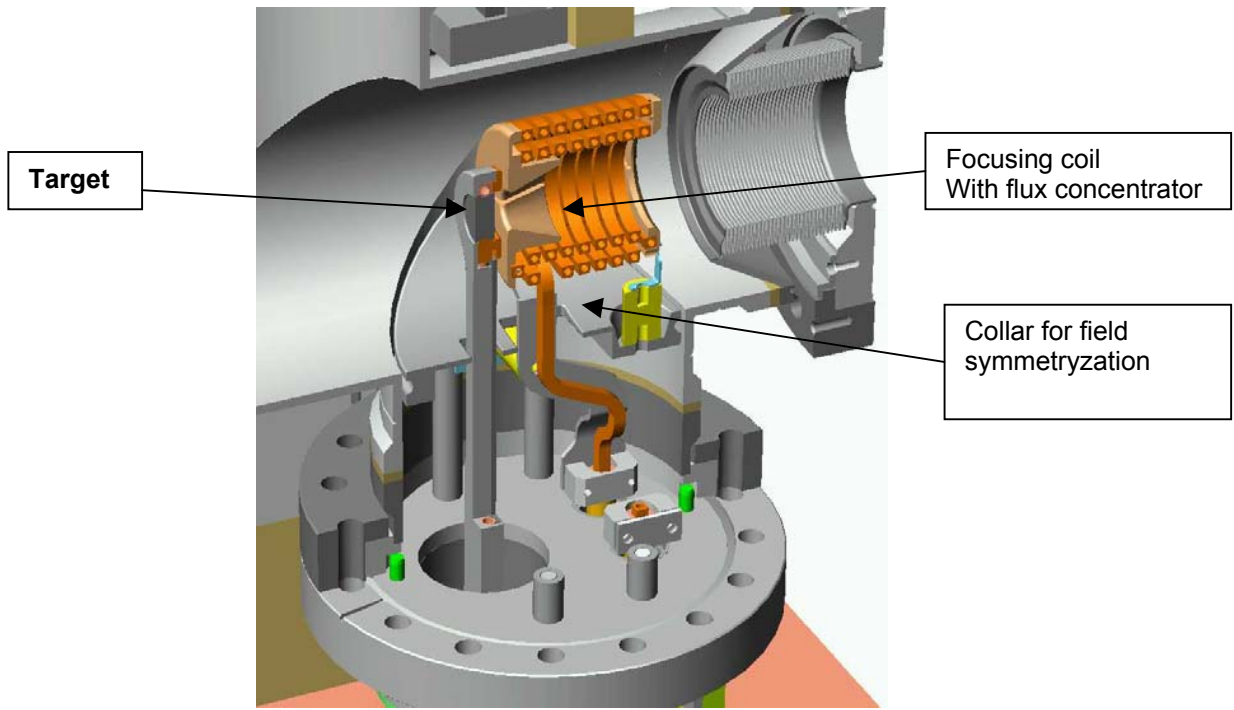

An isometric view of the focusing coil. The target is fixed at the end of the paddle-type holder.

The housing is made of Aluminum. This drastically reduces the losses from imaginary currents induced in the walls. Calculations show that the field reduction inside the coil due to surrounding walls is about 15% for the size of the tubing.

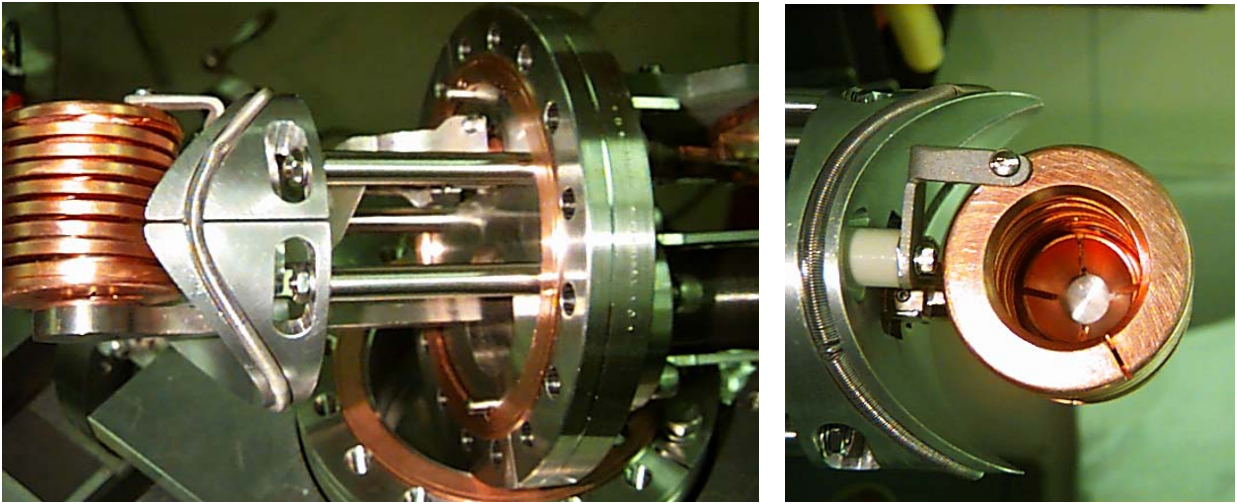

Photo of the focusing coil and the target.

Special collar like looking detail makes a continuation of the cylindrical shape around the coil. A helical spring makes a good contact between the part and the rest of the cabinet. Aluminum also reduces accumulation of isotopes.

*Measurements* were carried out with the small coil and integrator.

For calculation of magnetic field the model was used, which takes into account the skin depth of the current in flux concentrator media as well as in the target. For this the imaginary currents were applied running on the surfaces of the flux concentrator and target.

After that the currents found are positioned in PARMELA input file.

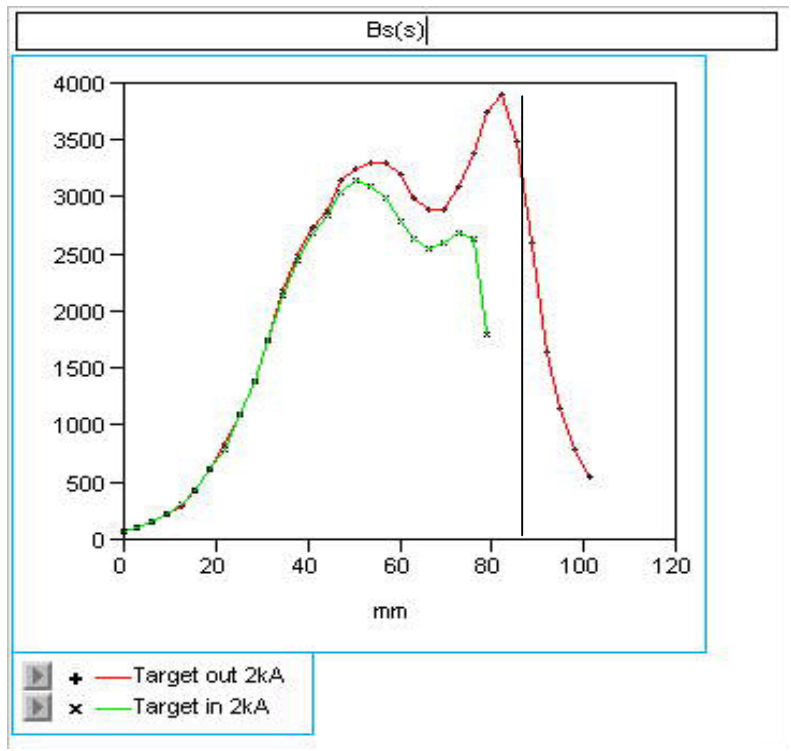

Longitudinal component of the field as a function of the distance, @2 kA. Peak field here is $\cong$ 10 kG. At the surface of the target this component ~zero. Azimuthal component, indeed, increases. Surface of the target marked by the line. Primary electron beam is coming from the right side of the picture.

The field at maximal point in Fig. is ~10*kG*, which is in good agreement with calculation.

**Allowed motion** is $\cong \pm 5$ *mm* in each direction.

**Strip line** made on copper sheets having width 4.2 *in* of outer two plates and 3.75 *in* central plate. Two layers of 10 *mils* Kapton insulate these electrodes. Line still flexible enough allowing motion of the cabinet. Closer to the cabinet the strip-line made with ten coaxial cables in parallel for flexible attachment to the pulser.

Paddle type holder of the target allows transverse motion of the target for removing it from the beam line during electron injection.

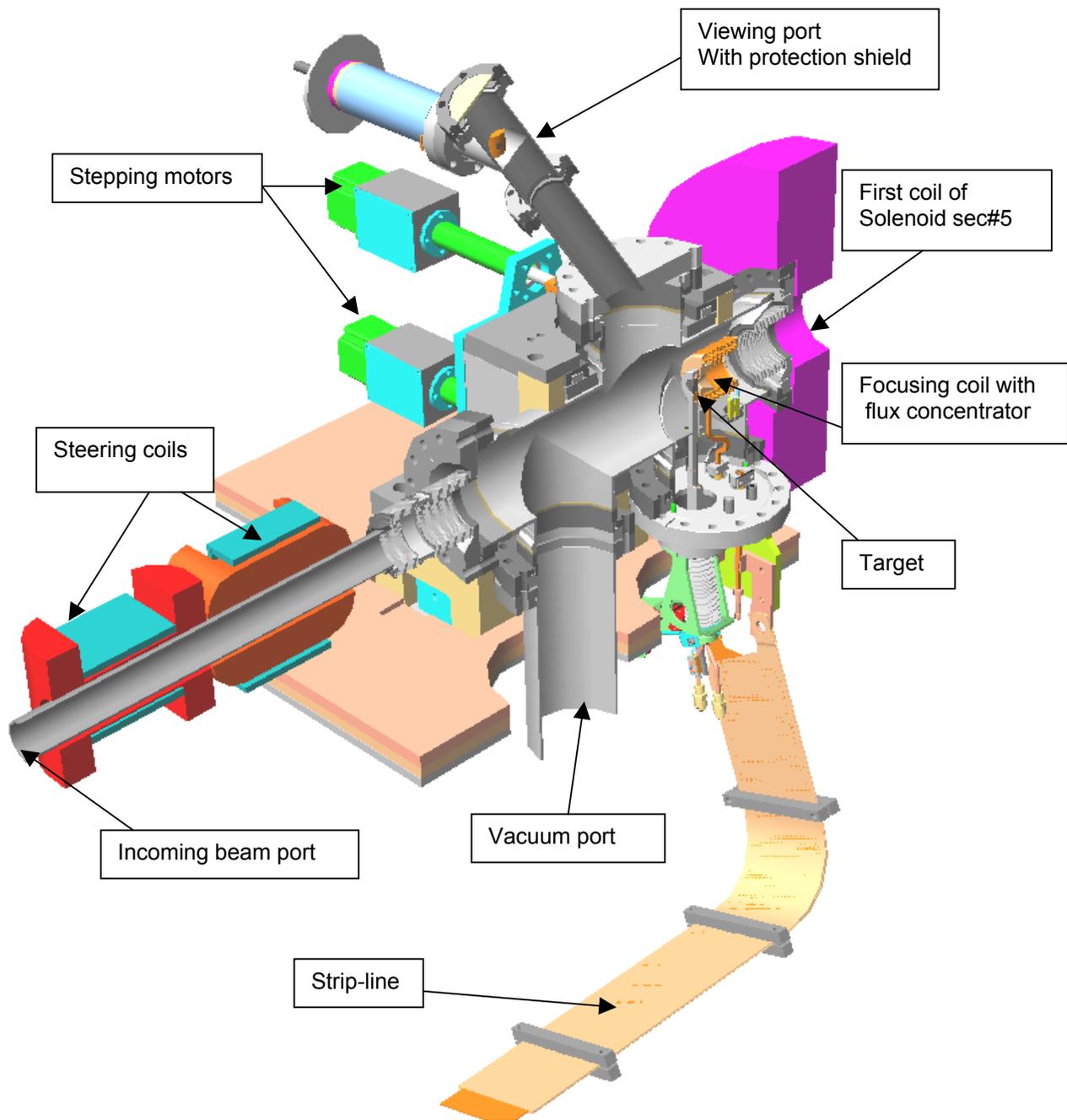

Converter unit. Focusing triplet located on incoming beam tube right before steering coils shown here and other viewing port (not shown in this figure).

# 4. PULSER UNIT

We implemented *modular* concept for power supply.

The pulser. Modular design allows easy replacement of triggering and thyristor modules.

We recognized importance of having large capacitor **C** and charging it with *constant current*. This procedure drastically reduces the losses allowing work with low charging voltage.

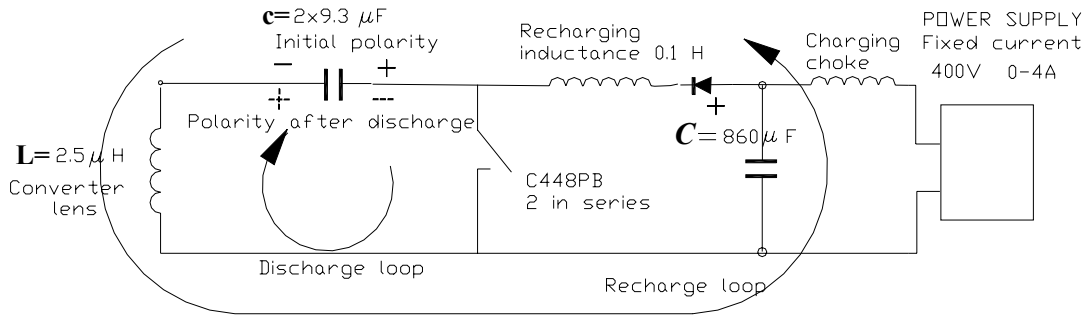

Equivalent scheme. The voltage hold on capacity of the power supply is adding to the recharging voltage every time during recharging. Power supply is charging capacity with *constant* (fixed) current.

Typically the voltage at capacitor **c** four times higher than that provided by power supply. For voltage on PS set to the level ~315 *V*, average current with this voltage goes to 4.4 *A* and pulsed current goes to ~4 *kA*.
The pulser tested for a long time period with modified lens at current ~4.6 *kA* and shortly for 5 *kA*.

Thyristor module was made to be easy removable.
Fast replacement time reduces the exposure time for personnel.

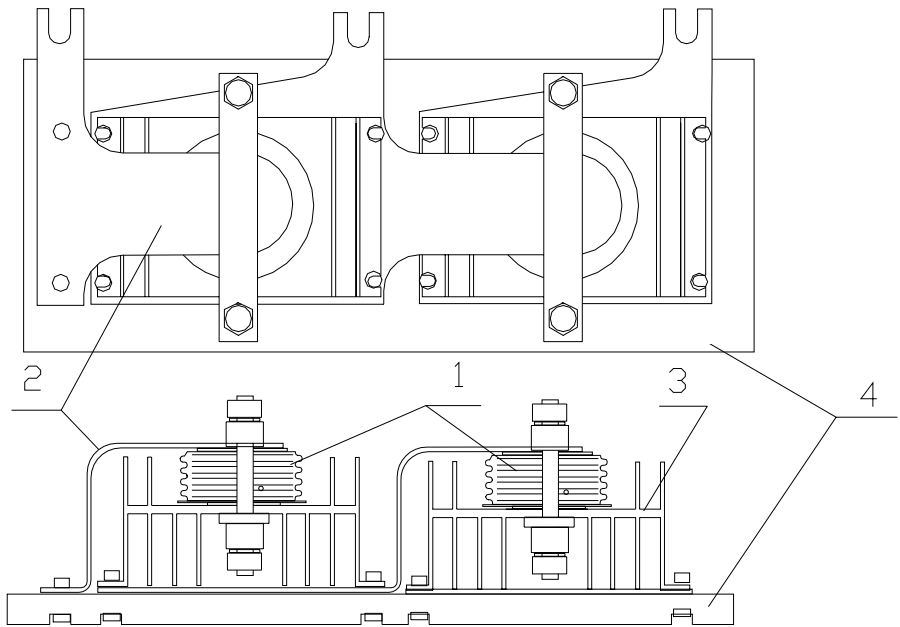

Thyristors' module. 1–thyristors, 2– the metallic plate-conductors, 3– a heat sink, 4– a G10 base.
.

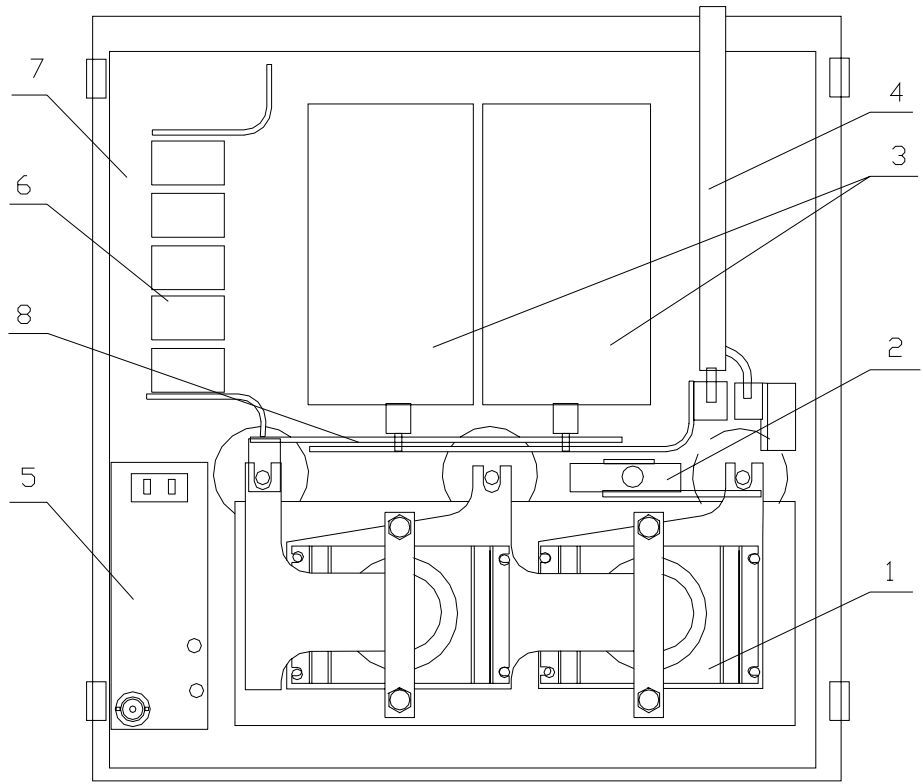

Structure scheme of the pulser. 1-the switch module from previous Figure. 3-capacitors *c*, 2-Rogowsky-type current monitor, 4-coaxial line, 5-triggering module, 6-diods, 7-metallic plate-basement, 8-wide inter conductors.

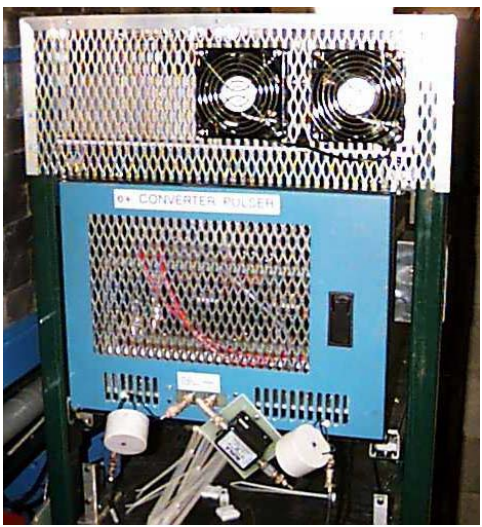 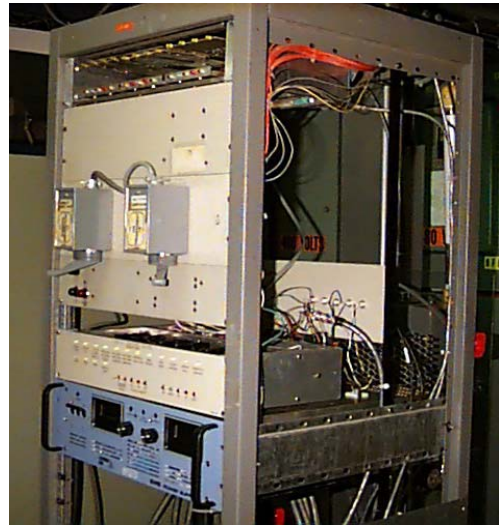

Pulser (left) and power supply with interlocks (right). RMS 600-8-2-D from LAMBDA EMI is used as a power supply.

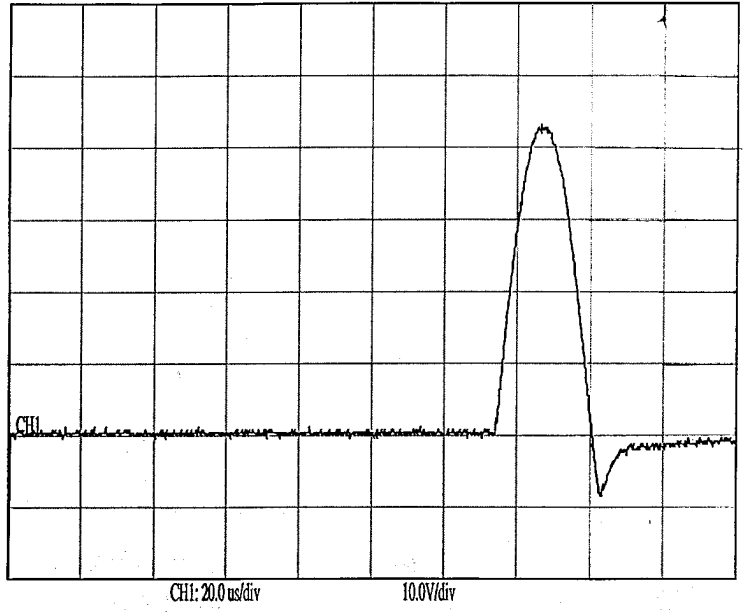

Signal from the current monitor read in control room. Horizontal scale 20 $\mu s$ /division. Vertical scale– 1kA/div.

Duty time over basement of the pulse is ~27 $\mu s$.

# 5. INSTALLATION INTO LINAC/TUNING RESULTS

Pulser installed closer to the target area in front of protection shield near section #4.

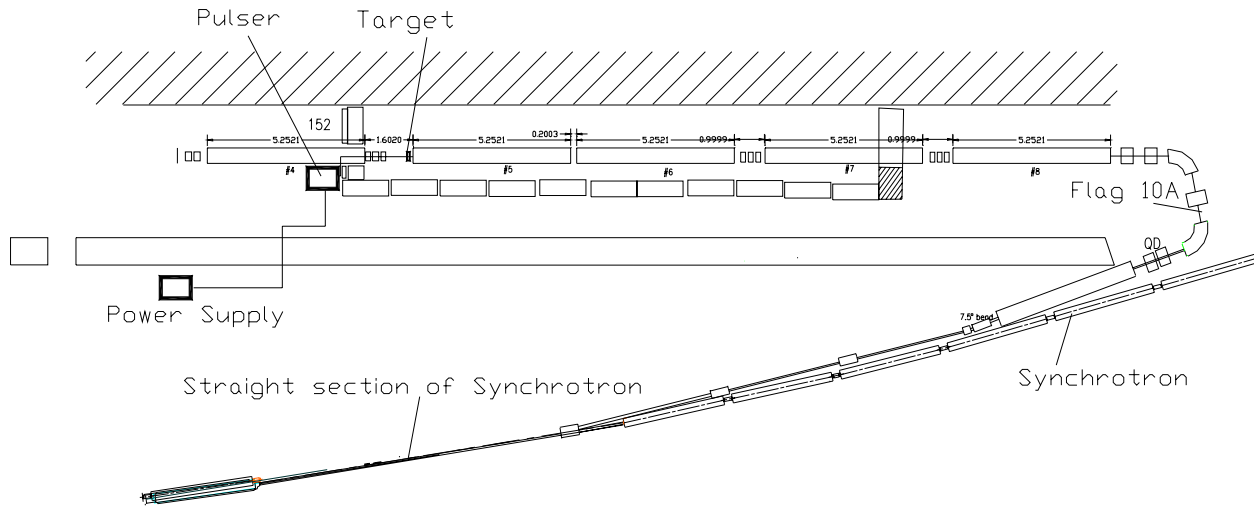

Positron channel map. Pulser installed near target area behind the concrete wall.

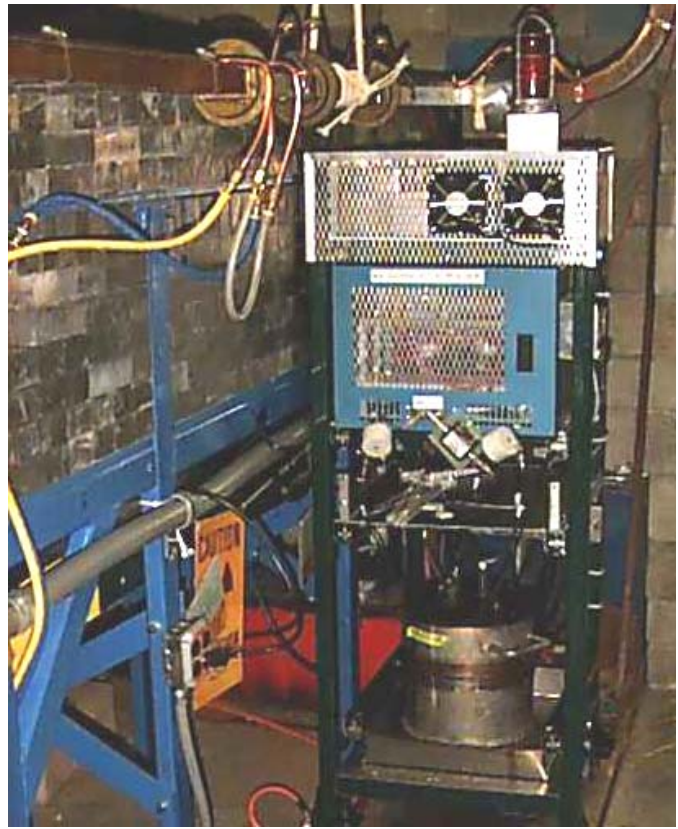

Pulser unit located in front of the converter cave. Section #4 is located at the left, behind the lead shield. Electron beam is running from the left side of the picture.

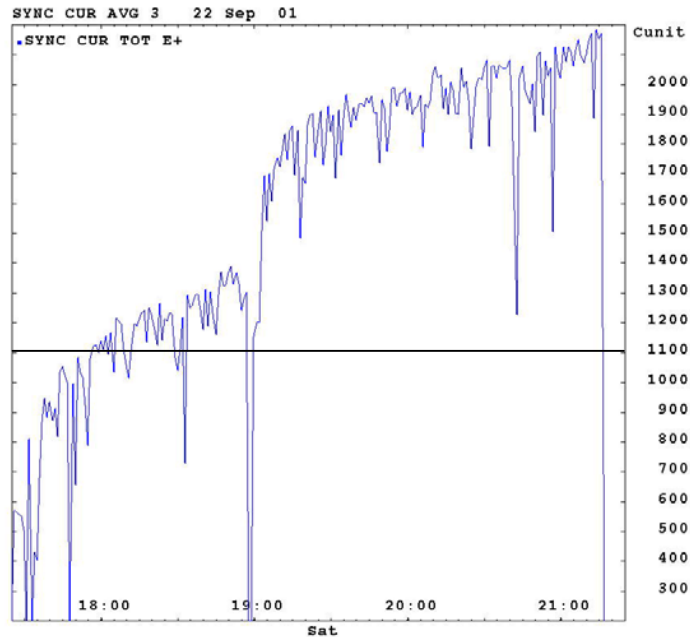

Tuning progress during the first run. Single bunch. Line marks previous achievements.

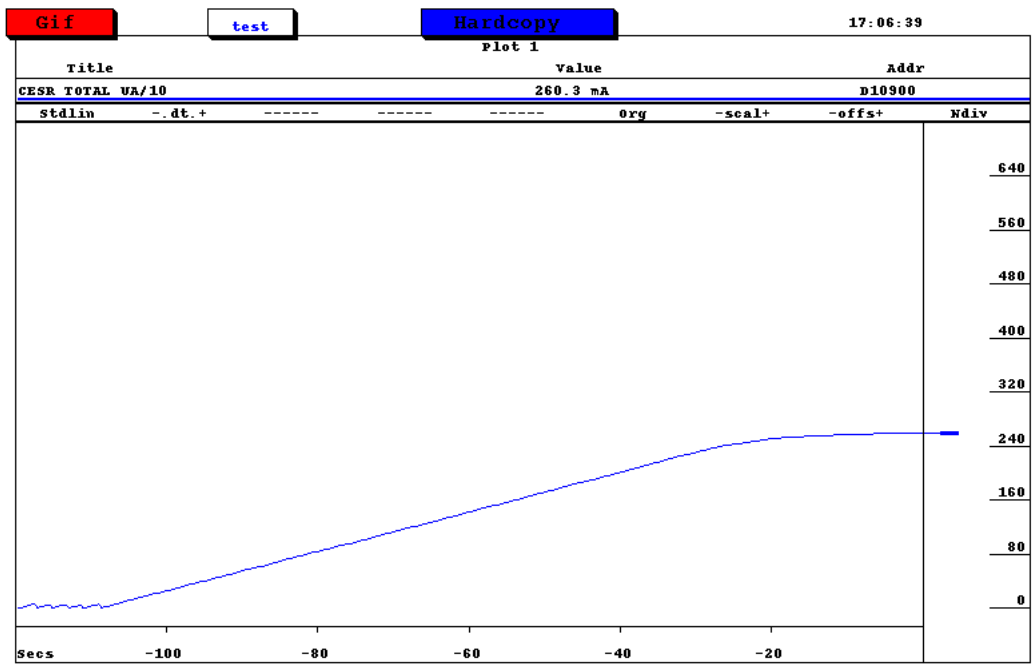

**Accumulation rate for positrons**

Right now even without any fine tuning the rate of accumulation of positrons in CESR is ~160*mA/ min* (Perimeter 768m).

```
2002-01-29 20:36:38.86 Tuesday (eve), entry 29

. . . . . . . . . Switched route:
Old Lattice =    CESRC_1S_V2         --> New = CESRC_1S_V2
Old TYPE:        ENCAL               New:     ENCAL
Old Energy,Sen1: 1S_+6   1797.78     New:     1S_+8   1798.54
Old Sen2, Dipole_status:  4950.40  1 New:     4952.50   1

Set      BIGGRP    POSCSR    ELCSR    ELTOLUM   POSTOP    COLLIDE   SCIR      HEP       CESR_RF
OLD      78200     78721     78733    78733     78733     78741     78722     78756     78237
NEW      78200     78723     78734    78734     78734     78742     78724     78479     78237

INJ Set  POSLIN    POSSYN    ELLIN    ELSYN
OLD      78713     78714     78715    78716
NEW      78713     78714     78715    78716

2002-01-29 20:57:31.27 Tuesday (eve), entry 30

Subject(s): ah dyptypsica
 after 3 energies without a hitch, lost beam when loaded "hep" set
 what a diff 78479 makes vs 78749
 (something not hep, not +8 meV from 22 jan...)
   now reslooping with right set.

 note that so far have had inj rates in all sets of 150 or more ma/min E+
 350-400 mA/min e-, and modest route loss in collide and hep, once
 pr 13 is set, and no other tuning (for inj or tau). as of now
 all + energy sets are ok to fill to 500 mA (6.66 ma e+, 0 nom advantage,
 .5+ mA actual e- advantage ) all sets so far like about 490 pr13 at 490 mA
 (as a mnemonic). -sbp -
  New ENCAL 1S_+8 HEP...old: 78479 -> new: 78749
savcom data

2002-01-29 21:12:20.51 Tuesday (eve), entry 31

Subject(s): route
   not  ever enough time, but  + sets are ok. - sbp -
  New ENCAL 1S_+8 HEP...old: 78749 -> new: 78757
savcom data

2002-01-29 21:12:38.44 Tuesday (eve), entry 32

. . . . . . . . . Switched route:
Old Lattice =    CESRC_1S_V2         --> New = CESRC_1S_V2
Old TYPE:        ENCAL               New:     ENCAL
Old Energy,Sen1: 1S +8   1798.54     New:     1S -4   1794.00
```

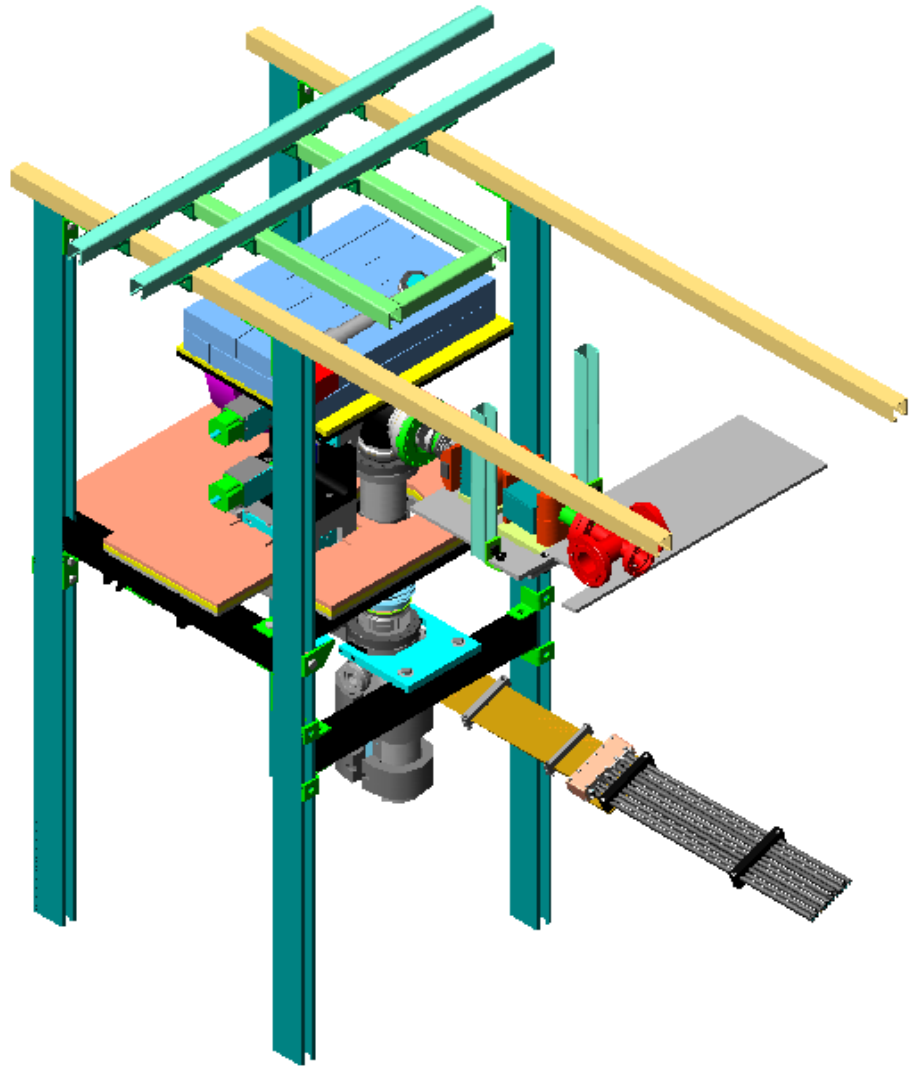

View of converter unit partially covered by led housing.

# 6. POSSIBLE IMPROVEMENTS

The points for further improvements are:
- Calculations carried with PARMELA code indicate that the efficiency can be increased up to 15% with first coil (#1 in first figure from chapter 3) off.
- Current can be risen up to 4.5 *kA* as the system was tested at this level. This might increase the capture up to 60%. Right now the pulsed current is kept at the level of 3.6 *kA* just for allowing CESR tunings without extremes.
- Possible installation of the focusing element(s) before the target. Right now the primary electron beam is focused on the target with the help of triplet assembled as a unit block. The midplane of this triplet located at $\cong 120 cm$ apart from the target. The steering coils mostly (one for each direction) occupy the space between the triplet and the converted housing.

Estimation to the maximal geometrical value of desirable beam size of the primary beam goes to 0.25 *mm*. Now this size is about 2.5*mm*. So there is evident necessity to decrease the size of the beam irradiating the target. This can increase the phase density of the positron beam as much as three times.

Limit for the material of target destruction under illumination by the bunch with population *N* can be taken from experimental work done at SLAC as [4][4]

$$NE/\pi\sigma^2 \cong 2 \cdot 10^{12} GeV/mm^2.$$

Here the *NE* is total energy carried by the bunch. The targets of optimal thickness supposed to be in use for every particular energy. We have $\sim 20 nC/pulse$ or $\sim 1.25 \cdot 10^{11}$ electrons, which yields $\sigma \geq 10^{-6}\sqrt{NE/2\pi} \cong 0.06 mm$. So the primary beam size can be reduced more than 40 times.

Additional focusing by Lithium lens for example can do this lowering of the primary beam spot size. The short focusing triplet can be also used here.

As *the cheapest solution* was recommend moving triplet closer to the target.

---

[4] [4] S. Ecklund, *Positron Target Materials Tests*, SLAC-CN-128, 1981.

# 7. CONCLUSION

Modifications of the positron converter unit done so far gave the basis for further increase of CESR's luminosity.
Collecting device can serve as a prototype for collection optics for future LC.

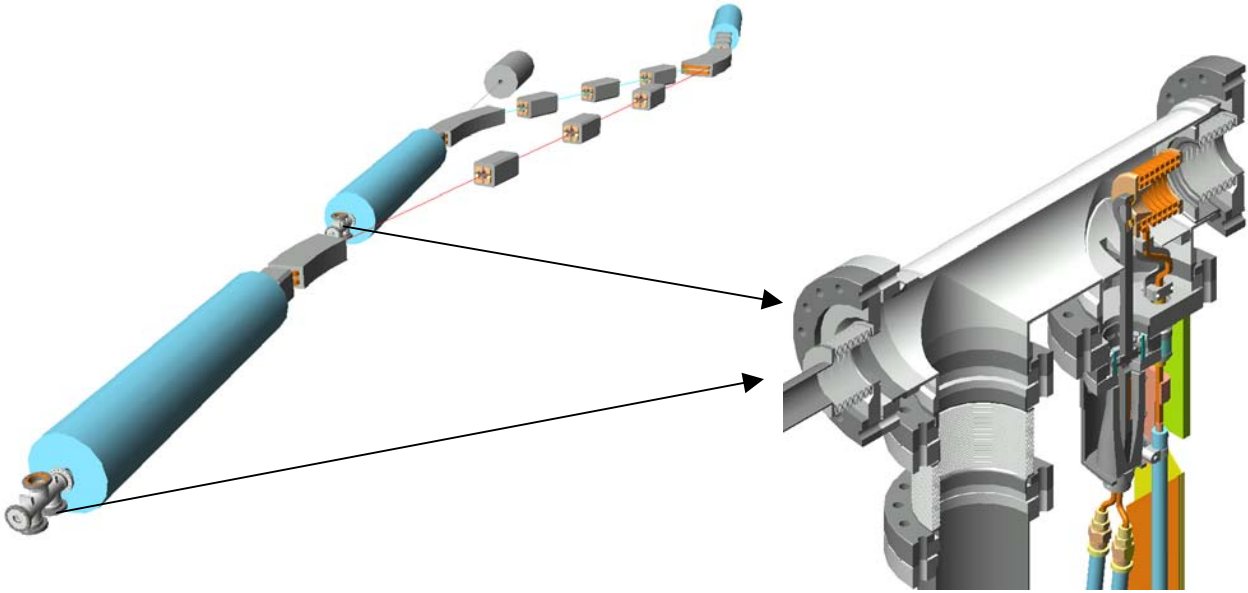

The solenoidal capturing optics of CESR's type used for positron conversion system using undulator.